\documentclass[10pt,conference]{IEEEtran}
\newif\ifextended
\extendedtrue

\usepackage{booktabs}
\usepackage{graphicx}
\usepackage{listings}
\usepackage{url}
\usepackage{xcolor}

\newcommand{\dsname}{CRUSTPairs}

\ifextended
  \newcommand{\runin}[1]{\textbf{#1}}
  \newcommand{\variant}[2]{#1}
\else
  \newcommand{\runin}[1]{#1}
  \newcommand{\variant}[2]{#2}
\fi

\begin{document}

\title{\dsname: A Cross-Language Dataset of Paired C and Rust Solutions}

\author{
\IEEEauthorblockN{Vojt\v{e}ch Hejlek}
\IEEEauthorblockA{Faculty of Information Technology in Prague\\
hejlevoj@fit.cvut.cz}
\and
\IEEEauthorblockN{Leonardo Jesus Almeida}
\IEEEauthorblockA{University of S\~ao Paulo\\
leonardo.j.almeida@usp.br}
\and
\IEEEauthorblockN{Alneu de Andrade Lopes}
\IEEEauthorblockA{University of S\~ao Paulo\\
alneu@icmc.usp.br}
}

\maketitle

\begin{abstract}
C-to-Rust migration is an active agenda, but no dataset pairs the two
languages at the level of whole solutions to a shared problem. We present
\dsname, 1857 C/Rust implementation pairs drawn from competitive programming
archives, normalised, deduplicated, and stratified into difficulty tiers by
zero-shot cross-language embedding similarity, with per-record provenance and
licensing. Cross-language retrieval is hard on this data, and harder than a
single baseline would suggest: TF-IDF over shared identifiers reaches
MRR $0.21$ and so beats a pretrained code model at $0.19$, which on the
hardest quarter of the data scores $1.7\times$ chance. Finetuning reaches
$0.79$, and lifts that hardest quarter $10\times$. Renaming every user
identifier settles what is being learned: the finetuned models keep 96--97\% of
their accuracy without them, while the pretrained model keeps 66\% --- less
than the lexical baseline does --- and on structurally divergent pairs the
finetuned models are \emph{better} without identifiers than with them.
\end{abstract}


\section{Introduction}
\label{sec:intro}

Memory-safety defects remain the dominant vulnerability class in C. Microsoft
has for two decades attributed roughly $70\%$ of its CVEs to that
class,\footnote{Reported by Microsoft's Security Response Centre for roughly
two decades, and restated by its Azure CTO in a RustConf 2025 keynote.} and
Android shows the substitution works: its annual memory-safety vulnerabilities
fell from 223 in 2019 to under 50 in 2024 as Rust adoption grew, with roughly
$1000\times$ lower defect density in Rust than in C and
C++.\footnote{``Rust in Android: Move Fast and Fix Things,'' Google Security
Blog, 2025.} Policy has followed, through CISA's memory-safe roadmaps and the
US cyber director's call to adopt memory-safe languages;\footnote{CISA, ``The
Case for Memory Safe Roadmaps,'' December 2023; Office of the National Cyber
Director, ``Back to the Building Blocks: A Path Toward Secure and Measurable
Software,'' February 2024.} DARPA's TRACTOR programme\footnote{``Eliminating
Memory Safety Vulnerabilities Once and For All,'' DARPA, July 2024.
\url{https://www.darpa.mil/news/2024/memory-safety-vulnerabilities}} funds the
automated translation of legacy C
to Rust outright, on the premise that the codebases that matter are too large
to port by hand.

Translation tools are consequently proliferating, and they are evaluated almost
entirely on whether their output compiles and passes the original test
suite~\cite{nitin2026c2saferrust,shetty2024syzygy}. That is the right primary
criterion. It is also incomplete --- a translation can pass every test and
still be an unfaithful rendering of its source --- and it is unavailable
precisely where it is most needed, since legacy C is what most often lacks
tests. The complementary question, \emph{are these two programs, in two
languages, the same program?}, is a similarity judgement rather than an
execution one, and answering it with a learned model requires paired
supervision.

No such corpus exists for this language pair. The parallel code datasets either
span many languages at snippet
granularity~\cite{zhu2022xlcost,zhu2022cost} or go deep on a single pair that
is not this one~\cite{ahmad2023avatar}, and the largest collections are
grouped by problem rather than aligned across
languages~\cite{puri2021codenet,lu2021codexglue}. None of them includes Rust.

We present \dsname, 1857 pairs of C and Rust implementations of the same
competitive programming problem, each with its statement, normalised entry
points, per-record provenance and licence, a difficulty tier derived from
zero-shot cross-language embedding similarity, and a frozen stratified split.

\variant{Measuring on it reframes what the task requires. A lexical
baseline---TF-IDF over shared identifiers, no model---beats}{A lexical
baseline beats} a pretrained code model zero-shot,
because competitive programmers reuse variable names across languages. Renaming
every user identifier separates the regimes: the pretrained model loses a third
of its accuracy, more than the lexical baseline does, while finetuned models
lose almost nothing.\variant{ Cross-language matching here is learnable, but
not from identifiers.}{}

\noindent\textbf{Contributions.}
\begin{itemize}
  \item \dsname, 1857 C/Rust solution pairs with problem statements,
        normalised entry points, per-record provenance and licensing.
  \item A difficulty stratification derived from zero-shot cross-language
        embedding similarity, independent of the model family used in
        evaluation, which we show transfers to a different family.
  \item Retrieval baselines per difficulty tier over a frozen stratified
        split, including a lexical baseline and an identifier ablation that
        together establish what the task does and does not reduce to.
\end{itemize}

\section{Related Datasets}
\label{sec:related}

Parallel code corpora divide by what they align and how many languages they
span. The largest are not parallel at all:
CodeNet~\cite{puri2021codenet} and POJ-104, distributed with
CodeXGLUE~\cite{lu2021codexglue}, group solutions by problem without
shipping an aligned bilingual corpus. Of the parallel ones,
xCodeEval~\cite{khan2024xcodeeval} is broadest at 17 languages;
XLCoST~\cite{zhu2022xlcost} and CoST~\cite{zhu2022cost} align snippets
rather than whole programs across the 7 languages GeeksForGeeks covers; and
AVATAR~\cite{ahmad2023avatar} goes deep on Java/Python instead. None includes
Rust, which is the gap \dsname{} fills and the reason it is one language pair
rather than a matrix.

\ifextended
\begin{table}[tbp]
\caption{Cross-language code corpora. \dsname{} is the only one pairing C with
Rust at solution granularity. Language counts are as reported by each corpus.}
\label{tab:related}
\centering
\begin{tabular}{lccc}
\toprule
Dataset & Languages & Paired & Rust \\
\midrule
CodeNet~\cite{puri2021codenet}         & 55  & no  & no \\
POJ-104~\cite{lu2021codexglue}         & 1   & n/a & no \\
xCodeEval~\cite{khan2024xcodeeval}     & 17  & yes & no \\
XLCoST~\cite{zhu2022xlcost}            & 7   & yes & no \\
CoST~\cite{zhu2022cost}                & 7   & yes & no \\
AVATAR~\cite{ahmad2023avatar}          & 2   & yes & no \\
\midrule
\dsname{}                              & 2   & yes & yes \\
\bottomrule
\end{tabular}
\end{table}
\fi

The modelling literature is correspondingly single-pair-poor. Unsupervised
program translation was established on C++/Java/Python by
TransCoder~\cite{roziere2020transcoder}, the pretrained encoders that
cross-language retrieval builds on---CodeBERT~\cite{feng2020codebert} and
UniXcoder~\cite{guo2022unixcoder}---saw little or no Rust in pretraining, and
cross-language clone detection has the same shape:
CLCDSA~\cite{nafi2019clcdsa} matches type-4 clones across Java, C\# and Python
with syntactic features and API-call similarity, stopping short of Rust.

\section{Construction}
\label{sec:construction}

\subsection{Sources and pairing}

\dsname{} draws on two competitive programming archives: xCodeEval
\cite{khan2024xcodeeval}, itself derived from Codeforces, and IBM Project
CodeNet \cite{puri2021codenet}, derived from AtCoder. Both distribute accepted
submissions, so the two implementations of a pair are known to have passed the
same official tests. A pair is one problem, one accepted C submission, and one
accepted Rust submission.

\subsection{Normalisation and cleaning}

Both sides pass through a static analysis pipeline that strips comments,
removes unused functions and macros, normalises encoding, and applies the
language's standard formatter. Entry points are renamed to \texttt{solution} in
both languages so the function name carries no signal---which also means the
distributed code no longer compiles standalone
(Section~\ref{sec:limitations}).

From an initial 2013 pairs, 127 were removed: 2 confirmed mis-pairs, 69 AtCoder
easy/hard variants with near-identical statements (description embedding
similarity $>0.90$), 54 further suspicious pairs ($0.80$--$0.90$) found the same
way with OpenAI \texttt{text-embedding-3-large}, 1 length-ratio outlier and 1
case of genuine implementation divergence. A further 29 textbook reference
implementations were dropped: they carried the only copyleft content and the
only records whose halves were under different licences
(Section~\ref{sec:availability}), and had no problem statement. This leaves
1857 pairs.

\subsection{Difficulty stratification}

Each pair is scored by the cosine similarity between zero-shot
\texttt{SFR-Embedding-Code-400M\_R} embeddings of its C and Rust sides, and
binned at the 25th and 75th percentiles into easy, medium and hard
(thresholds $0.7856$ and $0.8687$; similarity mean $0.826$, standard deviation
$0.058$). The raw score is distributed alongside the tier, so the cuts can be
redrawn without re-running the model. Pairs a
generic code model already embeds close together are near-literal translations;
pairs it separates diverge structurally and demand a real cross-language
remapping.

SFR rather than UniXcoder~\cite{guo2022unixcoder} on purpose: defining the
tiers with the family that is later finetuned would make hard pairs hard by
construction, and any per-tier improvement circular. The stratification is also
origin-agnostic, since AtCoder and Codeforces rate difficulty on incompatible
scales.

\section{Dataset Description}
\label{sec:description}

\subsection{Schema}

\dsname{} is a single 5.9\,MB JSON Lines file, one object per pair, with the
fields in Table~\ref{tab:schema}. Nothing is split across files and no field
needs to be joined from elsewhere.

\begin{table}[tbp]
\caption{Schema. One JSON object per line of \texttt{dataset.jsonl}; all
fields are strings except \texttt{difficulty\_score}.}
\label{tab:schema}
\centering
\footnotesize
\begin{tabular}{@{}l@{~~}p{4.3cm}@{}}
\toprule
Field & Values \\
\midrule
\texttt{problem\_id} & \texttt{0001}--\texttt{1886}, inherited \\
\texttt{origin} & \texttt{codenet} (839), \texttt{xcodeeval} (1018) \\
\texttt{license} & SPDX expression; follows \texttt{origin} \\
\texttt{problem\_description} & upstream statement, verbatim; \texttt{\_format} is
  \texttt{html} (839) or \texttt{text} (1018) \\
\texttt{c\_code}, \texttt{rust\_code} & entry points renamed \texttt{solution} \\
\texttt{difficulty} & \texttt{easy} (465), \texttt{medium} (928), \texttt{hard}
  (464); \texttt{\_score} is the raw similarity \\
\texttt{split} & \texttt{train} (1485), \texttt{val} (185), \texttt{test} (187) \\
\bottomrule
\end{tabular}
\end{table}

The fifth row is \texttt{problem\_description\_format}, abbreviated to
fit; \texttt{problem\_id} runs to 1886 over 1857 records, keeping an
earlier release's numbering. Two choices are worth naming. \texttt{license} is
per record, because the sources differ (Section~\ref{sec:availability}), so
filtering on \texttt{origin} is the only way to extract a commercially usable
subset. And \texttt{difficulty\_score} ships the raw similarity beside the
tier, so the quantile cuts can be redrawn without re-running the model.

\subsection{Statistics}

Solutions are short: a median of 516 characters of C and 749 of Rust, with
problem statements at a median of 1231. The Rust side runs about $1.5\times$
longer in characters (median ratio $1.50$) while being \emph{shorter} in lines
---31.4 against 33.6 on average---so Rust solutions pack more into each line.
A length-based heuristic therefore does not separate the two languages the way
the character counts alone would suggest.

\ifextended
\begin{table}[tbp]
\caption{Size of the two sides and of the problem statements, in characters.
The C maximum is a single 16460-character outlier; the next longest is 4649.}
\label{tab:sizes}
\centering
\begin{tabular}{lrrrr}
\toprule
 & Mean & Median & p90 & Max \\
\midrule
C, characters     & 616  & 516  & 1046 & 16460 \\
Rust, characters  & 851  & 749  & 1424 & 4888 \\
Statement, chars  & 1459 & 1231 & 2717 & 7368 \\
\midrule
C, lines          & 33.6 & & & \\
Rust, lines       & 31.4 & & & \\
\bottomrule
\end{tabular}
\end{table}

\begin{table}[tbp]
\caption{Composition by source and difficulty tier.}
\label{tab:composition}
\centering
\begin{tabular}{lrrrr}
\toprule
Source & Easy & Medium & Hard & Total \\
\midrule
CodeNet (AtCoder)      & 241 & 408 & 190 & 839 \\
xCodeEval (Codeforces) & 224 & 520 & 274 & 1018 \\
\midrule
Total                  & 465 & 928 & 464 & 1857 \\
\bottomrule
\end{tabular}
\end{table}
\fi

Source and difficulty are not independent. CodeNet contributes 241 easy, 408
medium and 190 hard pairs; xCodeEval 224, 520 and 274. The CodeNet portion
skews easy while
the xCodeEval portion skews hard. Reporting a per-tier result therefore
requires a split that controls for both.

\subsection{Splits}

The train/validation/test assignment is frozen in the data rather than
regenerated from a seed, and is stratified by source $\times$ difficulty at
80/10/10: 1485, 185 and 187 pairs. The test set mirrors the full dataset to
within 0.3 percentage points by source and 0.6 by tier.

\section{Baselines}
\label{sec:baselines}

Retrieval is C-to-Rust unless stated otherwise: each C implementation queries
the pool of all 187 Rust implementations in the test split, and the target is
the one solving the same problem.

We report mean reciprocal rank, introduced for the TREC-8 question answering
track~\cite{voorhees1999trec8} and standard for code search since
CodeSearchNet~\cite{husain2019codesearchnet} and CodeXGLUE~\cite{lu2021codexglue}:
each query contributes the reciprocal of the rank of its counterpart, so the
score turns on whether the answer lands at the top. MRR@10 truncates that,
contributing zero beyond rank ten; both are given because they diverge in the
zero-shot regime, which is the denominator of every ratio below. Chance is
$\mathrm{MRR}=0.031$ at this pool size. LoRA figures are mean $\pm$ sd over
four seeds; other rows are single runs.

\begin{table}[tbp]
\caption{Cross-language retrieval on the 187-pair test set.}
\label{tab:baselines}
\centering
\begin{tabular}{lccccc}
\toprule
Model & MRR & MRR@10 & R@1 & R@5 & R@10 \\
\midrule
Chance                 & 0.031 & 0.016 & 0.005 & 0.027 & 0.053 \\
Lexical TF-IDF         & 0.206 & 0.191 & 0.128 & 0.273 & 0.353 \\
\midrule
UniXcoder, zero-shot   & 0.189 & 0.170 & 0.134 & 0.203 & 0.283 \\
UniXcoder + LoRA       & 0.724 & 0.720 & 0.624 & 0.836 & 0.913 \\
UniXcoder, full        & \textbf{0.792} & \textbf{0.791} & \textbf{0.706} & \textbf{0.909} & \textbf{0.952} \\
\bottomrule
\end{tabular}
\end{table}

The lexical row is the one that fixes the frame. TF-IDF over shared
identifiers, with no model at all, reaches $0.206$ and so \emph{beats}
zero-shot UniXcoder~\cite{guo2022unixcoder} at $0.189$. Competitive programmers reuse variable names
across languages: \texttt{n} is shared by 85 of the 187 test pairs,
\texttt{i} by 84, then \texttt{a}, \texttt{t}, \texttt{s}, \texttt{j}. The
claim this dataset supports is therefore not that a generic code model finds
the task hard, but the stronger one that \emph{neither} generic code
embeddings \emph{nor} lexical matching solve it, and finetuning does ---
$3.8\times$ over zero-shot for LoRA~\cite{hu2022lora} and $4.2\times$ for
full finetuning
($4.2\times$ and $4.7\times$ on MRR@10, which diverges from MRR because many
zero-shot positives fall outside the top ten).

\subsection{Difficulty tiers and the identifier ablation}

Table~\ref{tab:bytier} reports MRR per tier, and repeats every configuration
with user identifiers renamed to positional names independently in each
language, which destroys cross-language name correspondence while leaving
control flow, types, library calls and string literals intact.

\begin{table}[tbp]
\caption{MRR by difficulty tier, with identifiers kept $\to$ renamed. Every tier is
scored against the same 187-candidate pool; only the queries are partitioned.
\emph{Ret.} is the anonymised overall MRR as a fraction of the normal one.}
\label{tab:bytier}
\centering
\begin{tabular}{@{}l@{~}cccc@{}}
\toprule
Model & Easy & Medium & Hard & Ret. \\
\midrule
Lexical TF-IDF       & .446$\to$.355 & .144$\to$.094 & .089$\to$.080 & 76\% \\
UniXcoder zero-shot  & .328$\to$.276 & .186$\to$.084 & .053$\to$.048 & 66\% \\
UniXcoder + LoRA     & .957$\to$.941 & .776$\to$.688 & .388$\to$\textbf{.429} & 96\% \\
UniXcoder, full      & .979$\to$.951 & .821$\to$.756 & .550$\to$\textbf{.611} & 97\% \\
\bottomrule
\end{tabular}
\end{table}

Three things follow, and together they are the strongest evidence that this
dataset measures what it is meant to.

\variant{\textbf{Retrieval is monotone in tier for every configuration.} The}{The} tiers were
cut with SFR while these models are UniXcoder-based, so the ordering is a
property of the pairs, not of the encoder that defined it. Zero-shot UniXcoder
reaches $0.053$ on the hard tier---$1.7\times$ chance, with one of the 47 hard
pairs at rank 1. A generic code model effectively cannot solve a quarter of
this dataset.

\variant{\textbf{Finetuning learns structure, not names.} The}{The} finetuned models keep
96--97\% of their MRR when identifiers are destroyed; zero-shot UniXcoder
keeps 66\%, \emph{less} than the lexical baseline's 76\%. A pretrained code
model is more name-dependent than TF-IDF, so it inverts the natural reading of
Table~\ref{tab:baselines}: zero-shot UniXcoder is not merely failing to beat
lexical matching---it is doing a worse version of the same thing.

\variant{\textbf{On hard pairs, identifiers are a liability.} Both}{Both} finetuned models do
\emph{better} without them---LoRA $0.388$ to $0.429$, full finetuning $0.550$ to
$0.611$, both about $+11\%$---while the lexical and zero-shot models get worse.
Where implementations diverge structurally, a shared name is more often
coincidence than signal.

\subsection{Direction and parameter efficiency}

Querying with Rust beats querying with C (MRR $0.189$ to $0.233$ zero-shot,
$0.792$ to $0.836$ full), and the asymmetry nearly vanishes under the ablation
($0.768$ against $0.773$), locating it in identifiers rather than structure.
LoRA reaches $91\%$ of full finetuning overall with $0.23\%$ of the parameters,
but $98\%$ on easy pairs against $71\%$ on hard: the budget binds where the
remapping is non-trivial, and the aggregate ratio conceals it.

\ifextended
\subsection{Training setup}

All runs finetune \texttt{microsoft/unixcoder-base} (126M parameters) with a
symmetric InfoNCE objective over in-batch negatives at temperature $0.07$:
for a batch of $N$ pairs the $N \times N$ cosine similarity matrix is computed
and each pair ranked against the other $N-1$, in both the C-to-Rust and
Rust-to-C directions. Embeddings are mean-pooled over non-padding tokens and
L2-normalised. Inputs are truncated at 512 tokens.

Batch size is 32, five epochs, with linear warmup over the first $10\%$ of
steps. Batch size is a property of the experiment rather than of the machine
here, because it sets how many negatives each step sees: a host with a larger
GPU running a larger batch would not produce a comparable number. It was
therefore held fixed across every run and every machine.

LoRA uses rank 8, $\alpha=16$, dropout $0.1$, applied to the query and value
projections of every attention layer---295K trainable parameters, $0.23\%$ of
the total---at learning rate $2 \times 10^{-4}$. Full finetuning updates all
126M at $1 \times 10^{-5}$ with weight decay $0.01$ and gradient
checkpointing. In both cases the checkpoint evaluated on test is the one with
the best validation MRR, and the LoRA runs use seeds 42 through 45.

The difficulty tiers come from \texttt{Salesforce/SFR-Em\-bedding-Code-400M\_R},
also mean-pooled and truncated at 512 tokens, with no finetuning.

\subsection{Full breakdowns}

Table~\ref{tab:bytierfull} gives every metric per tier rather than MRR alone,
and Table~\ref{tab:byorigin} the breakdown by source.

\begin{table}[tbp]
\caption{Test-split metrics by difficulty tier, identifiers kept. LoRA is the
mean over four seeds.}
\label{tab:bytierfull}
\centering
\footnotesize
\begin{tabular}{llcccc}
\toprule
Model & Tier & MRR & MRR@10 & R@1 & R@5 \\
\midrule
Lexical      & easy   & 0.446 & 0.439 & 0.319 & 0.596 \\
             & medium & 0.144 & 0.122 & 0.075 & 0.183 \\
             & hard   & 0.089 & 0.077 & 0.043 & 0.128 \\
\midrule
Zero-shot    & easy   & 0.328 & 0.313 & 0.234 & 0.383 \\
             & medium & 0.186 & 0.164 & 0.140 & 0.194 \\
             & hard   & 0.053 & 0.040 & 0.021 & 0.043 \\
\midrule
LoRA         & easy   & 0.957 & 0.957 & 0.915 & 1.000 \\
             & medium & 0.776 & 0.774 & 0.675 & 0.895 \\
             & hard   & 0.388 & 0.376 & 0.234 & 0.553 \\
\midrule
Full         & easy   & 0.979 & 0.979 & 0.957 & 1.000 \\
             & medium & 0.821 & 0.820 & 0.720 & 0.946 \\
             & hard   & 0.550 & 0.545 & 0.425 & 0.745 \\
\bottomrule
\end{tabular}
\end{table}

\begin{table}[tbp]
\caption{Test-split MRR by source, identifiers kept. The two portions have
slightly different tier mixes; reweighting accounts for $35\%$ of the LoRA gap
and $57\%$ of the smaller full-finetune one.}
\label{tab:byorigin}
\centering
\begin{tabular}{lcc}
\toprule
Model & CodeNet (84) & xCodeEval (103) \\
\midrule
Lexical TF-IDF       & 0.266 & 0.158 \\
UniXcoder, zero-shot & 0.217 & 0.165 \\
UniXcoder + LoRA     & 0.770 & 0.687 \\
UniXcoder, full      & 0.814 & 0.775 \\
\bottomrule
\end{tabular}
\end{table}

Source affects retrieval for the smaller model more than the larger one. Under
LoRA the gap between the two portions is $0.084$ and reweighting for their tier
mixes accounts for $35\%$ of it, so Codeforces problems are genuinely
harder to match. Full finetuning narrows it to $0.039$, of which composition
explains $57\%$: whatever distinguishes the sources is largely learnable given
enough capacity.
\fi

\section{Use Cases}
\label{sec:usecases}

Section~\ref{sec:intro} set out why C-to-Rust translation is being
funded rather than merely discussed. What follows is the gap between that
effort and what it currently measures.

\subsection{What a paired corpus contributes}

Translation tools are evaluated almost entirely on whether their output
compiles and passes the original tests; TRACTOR's own benchmark is 150 C
programs scored for correctness and
safety.\footnote{``Translating All C to Rust (TRACTOR) Benchmarks,'' MIT
Lincoln Laboratory. \url{https://www.ll.mit.edu/r-d/projects/}} That
is the right primary criterion and an incomplete one---a translation can pass
every test and still render its source unfaithfully---and it is unavailable
where it is most needed, since legacy C is what most often lacks tests. The
complementary question, \emph{are these two programs the same program?}, is a
similarity judgement, and it is the one \dsname{} supplies supervision for.

\variant{Two uses follow. \textbf{Scoring translator output without tests}: a}{A} model
finetuned on \dsname{} scores a (C, Rust) pair, flagging translations that
drift from their source, with the per-tier results bounding its usable
range---$0.55$ MRR on structurally divergent pairs, not the $0.98$ the easy
tier suggests. \variant{\textbf{Finding an existing port}: equivalent}{Equivalent} functionality
often already exists in a crate, and locating it is cross-language retrieval
over whole implementations.

\ifextended
A third follows from the first two. \textbf{Cross-language clone and
vulnerability tracking}: when a CVE is filed against a C implementation,
whether a Rust port inherits the defect---or whether a Rust crate is a
translation of vulnerable C---is a type-4 clone problem across languages.
Type-4 clone corpora are overwhelmingly single-language, and the cross-language
detectors that exist, such as CLCDSA~\cite{nafi2019clcdsa}, were evaluated on
language sets that stop short of Rust.

A fourth is more exploratory. Each pair is two solutions to one problem, one
written with a borrow checker and one without, which makes the dataset a
controlled sample for asking how ownership and lifetimes reshape an
implementation. The difficulty tiers already index this: the hard tier is
precisely the subset where the two languages' solutions diverge structurally
rather than translating line for line.
\fi

These are competitive programming solutions, without the allocation patterns,
aliasing and concurrency that make real C hard to port; read the results as a
measurement on a clean proxy rather than evidence about systems code.

\section{Limitations, Availability and Licensing}
\label{sec:limitations}
\label{sec:availability}

\runin{Equivalence is inherited, not verified.} No tests or expected I/O are
distributed; that both submissions were accepted upstream is evidence of
functional equivalence, not a check of it, and the entry-point renaming means
the code does not compile standalone. Execution-based verification is the
clearest future work. Relatedly, upstream identifiers are absent, so pairs
cannot be traced to the originating submission, and difficulty is a proxy for
embedding similarity rather than algorithmic difficulty. These are also
competitive programming solutions---short, single-file, I/O-driven---so nothing
here should be assumed to transfer to systems code, and pre-training
contamination is possible, both archives being public.

\variant{\textbf{Variance is measured for one configuration.} }{}The LoRA rows are means
over four seeds ($\sigma = 0.011$ on MRR); the others are single runs.
Separately, an earlier release's tier labels proved not exactly reproducible:
regenerating them matches the aggregate distribution but only 89\% of individual
labels, with disagreement spread symmetrically across adjacent tiers---which
threshold movement alone cannot produce. The released labels are the
regenerated ones, with raw scores alongside.

\ifextended

\textbf{On the reproducibility of the earlier labels, in detail.} The
disagreement was 204 of 1857 labels. Only one pair moved two tiers; the rest
moved one, and in both directions---57 medium-to-easy against 49
easy-to-medium, 51 hard-to-medium against 46 medium-to-hard. Dropping 29 pairs
shifted the quantile cuts slightly, but only 62 pairs lie between the old and
new thresholds, and threshold movement can only reassign in one direction.
Bidirectional movement at that volume therefore means the per-pair scores
differed between the runs, not merely the cuts. The cause is not recoverable,
because the original run stored bins and not scores; pooling variant,
truncation length, numeric precision and library version are all candidates.
For calibration, re-running the current script on a different machine---CPU
against GPU---moved 304 of 1857 scores by at most $1.2 \times 10^{-5}$ and
changed no label at all, so hardware non-determinism is not a plausible
explanation for the original discrepancy.
\fi

\dsname{}, the construction and evaluation scripts, the frozen split and the
trained adapter are archived at
\url{https://doi.org/10.5281/zenodo.23888785}, and developed at
\url{https://github.com/hejlevoj/crosslang-c-rust-similarity}.

\variant{\textbf{Licensing.} }{}\dsname{} aggregates two corpora under different
licences, so every record carries its own SPDX expression: the 1018
xCodeEval-derived pairs are CC BY-NC 4.0 and the 839 CodeNet-derived pairs are
CDLA-Permissive-2.0. The aggregate is therefore non-commercial and requires
attribution to both projects; the CodeNet portion alone gives 839 pairs with no
commercial restriction.

\bibliographystyle{IEEEtran}
\bibliography{refs}

\end{document}